\documentclass{ap-jnmp}

\copyrightauthor{}

\title{Integrable Boundary Conditions \\
for the Hirota-Miwa Equation and Lie Algebras}

\author{Ismagil Habibullin}

\address{Department of Mathematical Physics, Institute of Mathematics \\
Ufa Federal Research Centre, Russian Academy of Sciences\\ 
112, Chernyshevsky Street, Ufa 450008, Russian Federation\\
Bashkir State University, 32 Validy Street, Ufa 450076 , Russian Federation\\
\email{habibullinismagil@gmail.com}}

\author{Aigul Khakimova}

\address{Department of Mathematical Physics, Institute of Mathematics \\
Ufa Federal Research Centre, Russian Academy of Sciences\\ 
112, Chernyshevsky Street, Ufa 450008, Russian Federation\\
\email{aigul.khakimova@mail.ru}}

\begin{document}

\maketitle
\thispagestyle{empty}

\vphantom{\vbox{%
\begin{history}
\received{(Day Month Year)}
\revised{(Day Month Year)}
\accepted{(Day Month Year)}
\end{history}
}}

\begin{abstract}
Systems of discrete equations on a quadrilateral graph related to the series $D^{(2)}_N$ of the affine Lie algebras are studied. The systems are derived from the Hirota-Miwa equation by imposing boundary conditions compatible with the integrability property. The Lax pairs for the systems are presented. It is shown that in the continuum limit the quad systems tend to the corresponding systems of the differential equations belonging to the well-know Drinfeld-Sokolov hierarchies. The problem of finding the formal asymptotic expansion of the solutions to the Lax equations is studied. Generating functions for the local conservation laws are found for the systems corresponding to $D^{(2)}_3$. An example of the higher symmetry is presented.
\end{abstract}

\ccode{2000 Mathematics Subject Classification: 35Q51, 35Q58}

\section{Introduction}	
We consider the well-known Hirota-Miwa equation represented in the following form \cite{Hirota,Miwa}
\begin{equation}\label{HM}
at^j_{n,m}t^j_{n+1,m+1}-t^j_{n+1,m}t^j_{n,m+1}=bt^{j-1}_{n+1,m}t^{j+1}_{n,m+1}.
\end{equation}
Here $a$ and $b$ are constant parameters, the sought function $t^j_{n,m}$ depends on three integers $j,n,m$. Equation (\ref{HM}) is the most important discrete integrable equation in three dimensions. 
Years ago, bilinear equations of the form (\ref{HM}) have found applications \cite{Krichever,ZabrodinJETP} in the context of quantum integrable systems
as the model-independent functional relations for eigenvalues of quantum transfer matrices. 
Universality is a very remarkable property of this equation. As it was observed earlier by many authors (see \cite{ZabrodinTMF} and the references therein) numerous of known integrable continuous and discrete models can be derived from (\ref{HM}) by performing appropriate symmetry reductions, continuum limits etc. Moreover, it is generally accepted that almost all integrable models can be obtained from the Hirota-Miwa equation with the help of a suitable reduction. Initiated by this idea, we studied the problem of integrable boundary conditions for (\ref{HM}) in order to derive discrete versions of the integrable systems of exponential type, known as Drinfeld-Sokolov hierarchies \cite{Wilson,Drinfeld,Leznov79}.

We will interpret the equation (\ref{HM}) as an infinite sequence of related quadrilateral equations defined on the flat graph $(n, m)$ and endowed with the additional parameter $j$. Then we look for boundary  conditions that, when imposed at two selected points, say, $j=j_0$ and $j=j_{N}$ reduce the equation (\ref{HM}) to an integrable discrete system with a finite set of field variables $t^{j_0+1}_{n,m},t^{j_0+2}_{n,m},\cdots, t^{j_N-1}_{n,m}$.

Let us present some examples of quad systems connected with such kind reductions: 
\begin{eqnarray}
&&t^{-1}_{n,m}=1,\nonumber\\
&&at^j_{n,m}t^j_{n+1,m+1}-t^j_{n+1,m}t^j_{n,m+1}=bt^{j-1}_{n+1,m}t^{j+1}_{n,m+1},\quad 0\leq j\leq N-1,\quad\label{A-N}\\
&&t^{N}_{n,m}=1,\nonumber
\end{eqnarray}
\begin{eqnarray}
&&t^{-1}_{n,m}=1,\nonumber\\
&&at^j_{n,m}t^j_{n+1,m+1}-t^j_{n+1,m}t^j_{n,m+1}=bt^{j-1}_{n+1,m}t^{j+1}_{n,m+1},\quad 0\leq j\leq N-1,\quad\label{B-N}\\
&&t^{N}_{n,m+1}=t^{N-2}_{n+1,m},\nonumber
\end{eqnarray}
\begin{eqnarray}
&&t^{-1}_{n+1,m}=t^1_{n,m+1},\nonumber\\
&&at^j_{n,m}t^j_{n+1,m+1}-t^j_{n+1,m}t^j_{n,m+1}=bt^{j-1}_{n+1,m}t^{j+1}_{n,m+1},\quad 0\leq j\leq N-1,\quad\label{D2-N}\\
&&t^{N}_{n,m+1}=t^{N-2}_{n+1,m}.\nonumber
\end{eqnarray}
In the particular case when $a=1$, $b=1$ these systems were found in \cite{GHY}. S.V. Smirnov proved that for $a=b=1$ the quad systems (\ref{A-N}), (\ref{B-N}) are integrable in the sense of Darboux (see \cite{Smirnov}).

In our recent article \cite{HabKharxiv2019} we studied another reduction of this kind  
\begin{eqnarray}
&&t^{-1}_{n+1,m}=t^{N-1}_{n,m+1},\nonumber\\
&&at^j_{n,m}t^j_{n+1,m+1}-t^j_{n+1,m}t^j_{n,m+1}=bt^{j-1}_{n+1,m}t^{j+1}_{n,m+1},\quad 0\leq j\leq N-1,\quad\label{A1-N}\\
&&t^{N}_{n,m+1}=t^{0}_{n+1,m}.\nonumber
\end{eqnarray}
Note that for $N=2$ the systems (\ref{D2-N}) and (\ref{A1-N}) coincide with each other.

By introducing new variables $u^j_{n,m}=-\log t^j_{n,m}$ we can rewrite the equations (\ref{A-N})-(\ref{A1-N}) uniformly as follows
\begin{equation}\label{expon}
ae^{-u^i_{1,1}+u^i_{0,1}+u^i_{1,0}-u^i_{0,0}}-1=b\exp\left(\sum_{j=0}^{i-1}a_{i,j}u^j_{1,0}+\sum_{j=i+1}^{N-1}a_{i,j}u^j_{0,1}+\frac{a_{i,i}}{2}(u^i_{1,0}+u^i_{0,1})\right).
\end{equation}
Here $A=a_{i,j}$ is a constant matrix. For the cases (\ref{A-N}) and (\ref{B-N}) $A$ coincides with the Cartan matrices of the simple Lie algebras $A_{N-1}$ and $B_{N-1}$, respectively.
Similarly for (\ref{D2-N}) and (\ref{A1-N}) $A$ is the generalized Cartan matrix for the affine Lie algebra $D^{(2)}_N$ with $N\geq3$  and $A_{N-1}^{(1)}$ for $N\geq2$.

In our paper \cite{HabKharxiv2019} we derived the Lax pairs for the lattices (\ref{A1-N}) by imposing the quasi-periodical boundary condition $t^{j+N}_{n,m+1}=t^{j}_{n+1,m}$ on the Hirota-Miwa equation (\ref{HM}). It was observed in \cite{HabKharxiv2019} that this boundary condition generates a gluing condition for the Lax eigenfunctions $\psi^{j+N}_{n,m+1}=\lambda\psi^{j}_{n+1,m}$. This circumstance allowed us to simultaneously close both the nonlinear equation and its Lax pair. In \S2 of the present article we discuss one more boundary condition for (\ref{HM}) which is also compatible with the Lax representation. We observed that the boundary condition $t^{j-1}_{n+1,m}=t^{j+1}_{n,m+1}$ is consistent with a pair of the gluing conditions 
$$\psi^{j-1}_{n,m}=\lambda^{-1} g^j_{n,m+1} \quad\mbox{and}\quad g^{j-1}_{n,m}=\lambda \psi^j_{n-1,m}$$
connecting eigenfunctions of two essentially different Lax pairs for (\ref{HM}). These gluing conditions are crucial for deriving the Lax pair for the system (\ref{D2-N}).

The presence of the constant parameter $ b $ allows us to realize a continuum limit passage in the system (\ref{expon}). The limit is calculated in \S3. In the case of $D^{(2)}_N$, Lax pairs for quadrilateral systems tend to Lax pairs for the corresponding continuous systems. As an illustrative example, the system corresponding to $D^{(2)}_3 $ is considered.

One of the important applications of Lax pairs is the description of integrals of motion for the corresponding dynamical systems. Usually, for this purpose, asymptotic expansions of the Lax eigenfunctions around singular values of the spectral parameter are used. For discrete operators, the problem of constructing such expansions is complex and remains less studied. In \S4, we found transformations converting the Lax equations to a suitable form and constructed the necessary expansions, which, in turn, allowed us to construct generating functions for local conservation laws. This proves that the proposed Lax pairs are not ``fake''.

In the fifth section we presented a higher symmetry for the quad system corresponding to $D^{(2)}_3$. 

\section{Boundary conditions for the Hirota-Miwa equation}

The Hirota-Miwa equation (\ref{HM}) provides the consistency of the following overdetermined system of the linear equations
\begin{equation}\label{lax1}
\psi^j_{1,0}=\frac{t^{j+1}_{1,0}t^j_{0,0}}{t^{j+1}_{0,0}t^j_{1,0}}\psi^j_{0,0}-\psi^{j+1}_{0,0}, \quad
\psi^j_{0,1}=\psi^{j}_{0,0}+b\frac{t^{j+1}_{0,1}t^{j-1}_{0,0}}{t^j_{0,0}t^j_{0,1}}\psi^{j-1}_{0,0}
\end{equation}
which however doesn't define the Lax pair for (\ref{HM}) because the consistency of (\ref{lax1}) doesn't imply (\ref{HM}).

In our recent article \cite{HabKharxiv2019} we observed that the quasi-periodical constraint  $t^{j}_{n,m}=t^{j+N}_{n+1,m-1}$ and $\psi^{j+N}_{n,m}=\lambda \psi^{j}_{n+1,m-1}$ imposed on both variables simultaneously reduces the Hirota-Miwa equation (\ref{HM}) into an integrable quad system such that the linear system (\ref{lax1}) generates the Lax pair for this quad system. In other words the two constraints above are compatible. Below in this section we present one more example of a reduction of the equation (\ref{HM}) which is compatible with (\ref{lax1}).

Let us exclude $\psi^{j\pm1}$ from the system (\ref{lax1}) and arrive at a discrete equation of the hyperbolic type  
\begin{equation}\label{eqn1.1}
\psi^j_{1,1}-\psi^j_{1,0}-\frac{t^{j+1}_{1,1}t^j_{0,1}}{t^{j+1}_{0,1}t^j_{1,1}}\psi^j_{0,1}+
a\frac{t^j_{0,0}t^{j+1}_{1,1}}{t^j_{1,0}t^{j+1}_{0,1}}\psi^j_{0,0}=0
\end{equation}
for which the classical theory of the Laplace invariants can be applied \cite{AdlerStartsev,Novikov}. Recall that for a hyperbolic type linear equation 
\begin{equation}\label{f}
f_{1,1}+b_{0,0}f_{1,0}+c_{0,0}f_{0,1}+d_{0,0}f_{0,0}=0
\end{equation} 
the Laplace invariants are defined as follows
\begin{equation}\label{LaplaceInvar}
K_1=\frac{b_{0,0}c_{1,0}}{d_{1,0}},\quad 
K_2=\frac{b_{0,1}c_{0,0}}{d_{0,1}}.
\end{equation}
Equation (\ref{f}) and another equation of the same form 
\begin{equation}\label{tildef}
\tilde f_{1,1}+\tilde b_{0,0} \tilde f_{1,0}+\tilde c_{0,0}\tilde f_{0,1}+\tilde d_{0,0}\tilde f_{0,0}=0
\end{equation} 
are related by linear change of the variables $f=\lambda\tilde f$ if and only if their Laplace invariants coincide: $K_1=\tilde K_1$ and $K_2=\tilde K_2$.

Here to study the integrable finite-field reductions of the equation (\ref{HM}) we use the method of nonlinear mirror images (see \cite{Sklyanin, Habibullin1990, Habibullin2005, Biondini2009, Caudrelier2014, Caudrelier2018}). In order to use this method we need in addition to (\ref{lax1}) one more system of linear equations also associated with the Hirota-Miwa equation (\ref{HM})
\begin{equation}\label{lax3}
\begin{array}{l}
g^{j}_{-1,0}=\gamma^j
\left(g^{j}_{0,0}+g^{j-1}_{0,0}\right),\qquad
g^{j}_{0,-1}=\delta^jg^{j}_{0,0}-
\rho^jg^{j+1}_{0,0},
\end{array}
\end{equation}
where $\gamma^j=\frac{t^{j+1}_{-1,0}t^{j}_{-1,0}}{a t^{j}_{0,0}t^{j+1}_{-2,0}}$, $\delta^j=\frac{t^{j+1}_{0,-1}t^{j+1}_{-1,0}}{t^{j+1}_{0,0}t^{j+1}_{-1,-1}a}$ and 
$\rho^j=\frac{t^{j}_{0,-1}t^{j+2}_{-1,0}b}{t^{j+1}_{0,0}t^{j+1}_{-1,-1}a}$. It is easily checked that if the function $t^j_{n,m}$ solves (\ref{HM}) then system (\ref{lax3}) is consistent. It is important that the systems (\ref{lax1}), (\ref{lax3}) are not gauge equivalent.  In what follows we use also the discrete hyperbolic type equation
\begin{equation}\label{eqn3.1} 
g^j_{1,1}-g^j_{1,0}-\frac{t^j_{1,0}t^{j+1}_{-1,1}}{t^{j+1}_{0,1}t^j_{0,0}}g^j_{0,1}+
\frac{a t^j_{1,0}t^{j+1}_{-1,0}}{t^{j+1}_{0,0}t^j_{0,0}}g^j_{0,0}=0,
\end{equation}
which is a consequence of the equations (\ref{lax3}). Our goal is to construct the Lax pair for the quad system (\ref{D2-N}) by combining the equations (\ref{lax1}), (\ref{lax3}). We first study the question of when hyperbolic type equations (\ref {eqn1.1}) and (\ref {eqn3.1}) are connected by a multiplicative transformation.
To answer the question we have to compare the Laplace invariants of these equations. Denote through $K_{1\psi}(n,m,j)$ and $K_{2\psi}(n,m,j)$ the Laplace invariants for the equation (\ref {eqn1.1}) and respectively through $K_{1g}(n,m,j)$ and $K_{2g}(n,m,j)$ for the equation (\ref {eqn3.1}). Due to the formula (\ref{LaplaceInvar}) we have explicit representations
\begin{equation*}
K_{1\psi}=\frac{t^j_{2,0}t^j_{1,1}}{t^j_{1,0}t^j_{2,1}a},\quad K_{2\psi}=\frac{t^{j+1}_{1,1}t^{j+1}_{0,2}}{t^{j+1}_{0,1}t^{j+1}_{1,2}a},\quad
K_{1g}=\frac{t^{j+1}_{0,1}t^{j+1}_{1,0}}{t^{j+1}_{0,0}t^{j+1}_{1,1}a},\quad
K_{2g}=\frac{t^j_{1,0}t^j_{0,1}}{t^j_{0,0}t^j_{1,1}a}.
\end{equation*}
Evidently we have coincidence of the first invariants $K_{1\psi}(n,m,j)=K_{1g}(n+1,m,j-1)$ without any additional assumption on the function $t^j=t^j(n,m)$. Now if we assume the coincidence also of the second invariants $K_{2\psi}(n,m,j)=K_{2g}(n+1,m,j-1)$ then we obtain the equation
$$\frac{t^{j+1}_{1,1}t^{j+1}_{0,2}}{t^{j+1}_{0,1}t^{j+1}_{1,2}}=\frac{t^{j-1}_{2,0}t^{j-1}_{1,1}}{t^{j-1}_{1,0}t^{j-1}_{2,1}}$$
which is easily solved
\begin{equation*}\label{variable}
t^{j-1}_{1,0}=\hat a(n)\hat b(m) t^{j+1}_{0,1}.
\end{equation*}
Here $ \hat a (n) $ and $ \hat b (m) $ are arbitrary functions different from zero. However, the freedom in choosing of the factors is deceptive, since they are eliminated by an appropriate point transformation of the restricted system. Therefore it is reasonable to focus on such a choice
\begin{equation}\label{constant}
t^{j-1}_{1,0}=t^{j+1}_{0,1}.
\end{equation}

In a manner similar to that applied in \cite{HabKharxiv2019} we can show that the constraint (\ref{constant}) generates the gluing conditions of the form
\begin{equation}\label{gluing2}
\psi^{j-1}_{0,0}=\lambda^{-1} g^j_{0,1}, \quad g^{j-1}_{0,0}=\lambda \psi^j_{-1,0},
\end{equation}
where $\lambda$ is an arbitrary constant.

Presence of the gluing conditions indicates the consistency of the boundary condition with the integrability property of the equation (\ref{HM}). Now we are ready to construct the Lax pair for (\ref{D2-N}). Let us impose boundary conditions of the form (\ref{constant}) at the endpoints $N_L$ and $N_R$ of the segment $[N_L,N_R]$:
\begin{equation}\label{BC}
t^{N_L-1}_{1,0}=t^{N_L+1}_{0,1},\quad t^{N_R-1}_{1,0}=t^{N_R+1}_{0,1}.
\end{equation}
Due to the reasonings above they generate two pairs of gluing conditions
\begin{equation}\label{gluingL}
\psi^{N_L-1}_{n,m}=\lambda^{-1} g^{N_L}_{n,m+1}, \quad g^{N_L-1}_{n,m}=\lambda \psi^{N_L}_{n-1,m},
\end{equation}
\begin{equation}\label{gluingR}
g^{N_R}_{n,m}=\psi^{N_R-1}_{n,m-1}, \quad \psi^{N_R}_{n,m}=g^{N_R-1}_{n+1,m}.
\end{equation}
The gluing conditions allow immediately to derive a finite closed subsystem of the combined system (\ref{lax1}), (\ref{lax3}) for the eigenfunctions $\psi^j$, $g^j$ with $j\in[N_L,N_R-1]$, indeed we have
\begin{equation}\label{2211}
\psi^j_{1,0}=\alpha^j\psi^j_{0,0}-\psi^{j+1}_{0,0}, \quad N_L\leq j\leq N_R-2,
\end{equation}
\begin{equation}\label{2212}
\psi^{N_R-1}_{1,0}=\alpha^{N_R-1}\psi^{N_R-1}_{0,0}-g_{1,0}^{N_R-1}, 
\end{equation}
\begin{equation}\label{2213}
\psi^{N_L}_{0,1}=\psi^{N_L}_{0,0}+\beta^{N_L}\lambda^{-1} g_{0,1}^{N_L}, 
\end{equation}
\begin{equation}\label{2214}
\psi^j_{0,1}=\psi^j_{0,0}+\beta^j\psi^{j-1}_{0,0}, \quad N_L+1\leq j\leq N_R-1,
\end{equation}
\begin{equation}\label{2215}
g^{N_L}_{0,0}=\gamma^{N_L}_{1,0}(g^{N_L}_{1,0}+\lambda\psi^{N_L}_{0,0}), 
\end{equation}
\begin{equation}\label{2216}
g^{j}_{0,0}=\gamma^j_{1,0}(g^{j}_{1,0}+g^{j-1}_{1,0}), \quad N_L+1\leq j\leq N_R-1,
\end{equation}
\begin{equation}\label{2217}
g^{j}_{0,0}=\delta^j_{0,1}g^{j}_{0,1}-\rho^{j}_{0,1}g^{j+1}_{0,1}, \quad N_L\leq j\leq N_R-2,
\end{equation}
\begin{equation}\label{2218}
g^{N_R-1}_{0,0}=\delta^{N_R-1}_{0,1}g^{{N_R-1}}_{0,1}-\rho^{{N_R-1}}_{0,1}\psi^{{N_R-1}}_{0,0}, 
\end{equation}
where 
\begin{equation*}
\alpha^j=\frac{t_{1,0}^{j+1}t^j_{0,0}}{t^{j+1}_{0,0}t^j_{1,0}},\quad \beta^j=b\frac{t^{j+1}_{0,1}t^{j-1}_{0,0}}{t^{j}_{0,1}t^j_{0,0}}.
\end{equation*}
The obtained system of the equations (\ref{2211})-(\ref{2218}) can be rewritten in a compact form
$$A\Phi_{1,0}=B\Phi, \quad R\Phi_{0,1}=S\Phi,$$
where $\Phi$ is a column-vector $\Phi=(\psi^{N_L},\psi^{N_L+1},...,\psi^{N_R-1},g^{N_L},g^{N_L+1},...,g^{N_R-1})^T$ and $A$, $B$, $R$, $S$ are matrices. Finding inverse matrices, we can present the system in the usual form:
\begin{equation*}\label{matrix}
 \Phi_{1,0}=A^{-1}B\Phi, \quad \Phi_{0,1}=R^{-1}S\Phi.
\end{equation*}

However the more convenient way is to express function $g^j_{1,0}$ from equation (\ref{2216}) consecutively  by using equation (\ref{2215}) for determining $g^{N_L}_{1,0}$. 
As a result we obtain
\begin{equation}
g^{j}_{1,0}=\sum\limits^{j}_{k=N_L}(-1)^{j-k}\frac{at^k_{1,0}t^{k+1}_{-1,0}}{t^k_{0,0}t^{k+1}_{0,0}}g^k_{0,0}+(-1)^{j+1-N_L}\lambda\psi^{N_L}_{0,0}, \label{gLax_jn}
\end{equation}
where $N_L \leq j \leq N_R-1$.
Similarly we can express $g^{j+1}_{0,1}$ from (\ref{2217}) and use (\ref{2218}) for finding $g^{N_R-1}_{0,1}$
\begin{equation}
g^{j}_{0,1}=\frac{1}{\delta^j_{0,1}}g^{j}_{0,0}
+\sum\limits^{N_R-1}_{k=j+1}ab^{k-j}\frac{t^j_{0,0}t^{k+1}_{-1,0}t^{k+1}_{0,1}}{t^{j+1}_{-1,1}t^{k}_{0,0}t^{k+1}_{0,0}}g^k_{0,0}+
b^{N_R-j}\frac{t^j_{0,0}t^{N_R-1}_{0,0}}{t^{j+1}_{-1,1}t^{N_R}_{0,0}}\psi^{N_R-1}_{0,0}, \label{gLax_jm}
\end{equation}
where  $N_L \leq j \leq N_R-1$.
Now by combining (\ref{2211})-(\ref{2218}), (\ref{gLax_jn}) and (\ref{gLax_jm}) we find the final form of the Lax pair for the quad system (\ref{D2-N}):
\noindent
\begin{eqnarray}
&&\psi^{j}_{1,0}=\frac{t^{j+1}_{1,0}t^{j}_{0,0}}{t^{j+1}_{0,0}t^{j}_{1,0}}\psi^{j}_{0,0}-\psi^{j+1}_{0,0},\quad N_L\leq j\leq N_R-2,\label{psi_NL}\\
&&\psi^{N_R-1}_{1,0}=\lambda(-1)^{N_R-N_L-1}\psi^{N_L}_{0,0}  +\frac{t^{N_R-1}_{0,0}t^{N_R}_{1,0}}{t^{N_R-1}_{1,0}t^{N_R}_{0,0}}\psi^{N_R-1}_{0,0}+\sum\limits^{N_R-1}_{k=N_L}(-1)^{N_R-k}\frac{at^k_{1,0}t^{k+1}_{-1,0}}{t^k_{0,0}t^{k+1}_{0,0}}g^k_{0,0},
 \label{psi_N_Rn}\\
&&g^{j}_{1,0}=\sum\limits^{j}_{k=N_L}(-1)^{j-k}\frac{at^k_{1,0}t^{k+1}_{-1,0}}{t^k_{0,0}t^{k+1}_{0,0}}g^k_{0,0}+(-1)^{j+1-N_L}\lambda\psi^{N_L}_{0,0}, \quad N_L \leq j \leq N_R-1,\label{gLax_jn2}\\
&&\psi^{N_L}_{0,1}=\psi^{N_L}_{0,0}+ \lambda^{-1} b^{N_R-N_L+1}\frac{t^{N_L+1}_{0,1}t^{N_R-1}_{0,0}}{t^{N_L}_{0,1}t^{N_R}_{0,0}}\psi^{N_R-1}_{0,0}+\lambda^{-1} \sum\limits^{N_R-1}_{k=N_L}ab^{k-N_L+1}
\frac{t^{N_L+1}_{0,1}t^{k+1}_{0,1}t^{k+1}_{-1,0}}{t^{N_L}_{0,1}t^k_{0,0}t^{k+1}_{0,0}}g^k_{0,0},\label{psi_N_Lm}\\
&&\psi^{j}_{0,1}=\psi^{j}_{0,0}+\frac{bt^{j+1}_{0,1}t^{j-1}_{0,0}}{t^{j}_{0,0}t^{j}_{0,1}}\psi^{j-1}_{0,0},\quad N_L+1\leq j\leq N_R-1,\\
&&g^{j}_{0,1}=\frac{at^{j+1}_{-1,0}t^{j+1}_{0,1}}{t^{j+1}_{0,0}t^{j+1}_{-1,1}}g^{j}_{0,0}
+\sum\limits^{N_R-1}_{k=j+1}ab^{k-j}\frac{t^j_{0,0}t^{k+1}_{-1,0}t^{k+1}_{0,1}}{t^{j+1}_{-1,1}t^{k}_{0,0}t^{k+1}_{0,0}}g^k_{0,0}+\nonumber\\
&&\quad \quad b^{N_R-j}\frac{t^j_{0,0}t^{N_R-1}_{0,0}}{t^{j+1}_{-1,1}t^{N_R}_{0,0}}\psi^{N_R-1}_{0,0}, \quad N_L \leq j \leq N_R-1. \label{gLax_jm2}
\end{eqnarray}
We rewrite the Lax pair (\ref{psi_NL})-(\ref{gLax_jm2}) of system (\ref{D2-N}) under the condition $N_L=0$ and $N_R=N$ as
\begin{equation}\label{LaxD2N1}
 \Phi_{1,0}=F\Phi, \quad \Phi_{0,1}=G\Phi,
\end{equation}
where  $\Phi=(\psi^{0},\psi^{1},...,\psi^{N-1},g^{0},g^{1},...,g^{N-1})^T$,
\begin{equation*} \label{DNf}
F=\left(\begin{array}{cccccccc}
\frac{t^{1}_{1,0}t^{0}_{0,0}}{t^{1}_{0,0}t^{0}_{1,0}}&-1&0&\dots &0&0&\dots &0 \\
0&\frac{t^{2}_{1,0}t^{1}_{0,0}}{t^{2}_{0,0}t^{1}_{1,0}}&-1&\dots &0&0&\dots &0 \\
\dots &\dots&\dots& \dots&\dots&\dots&\dots&\dots \\
(-1)^{N-1}\lambda&0&\dots&\frac{t^{N}_{1,0}t^{N-1}_{0,0}}{t^{N}_{0,0}t^{N-1}_{1,0}} &(-1)^{N}\frac{at^{0}_{1,0}t^{1}_{-1,0}}{t^{0}_{0,0}t^{1}_{0,0}}&(-1)^{N-1}\frac{at^{1}_{1,0}t^{2}_{-1,0}}{t^{1}_{0,0}t^{2}_{0,0}}&\dots &-\frac{at^{N-1}_{1,0}t^{N}_{-1,0}}{t^{N-1}_{0,0}t^{N}_{0,0}} \\
-\lambda&0&\dots&0 &\frac{at^{0}_{1,0}t^{1}_{-1,0}}{t^{0}_{0,0}t^{1}_{0,0}}&0&\dots &0 \\
\lambda&0&\dots&0 &-\frac{at^{0}_{1,0}t^{1}_{-1,0}}{t^{0}_{0,0}t^{1}_{0,0}}&\frac{at^{1}_{1,0}t^{2}_{-1,0}}{t^{1}_{0,0}t^{2}_{0,0}}&\dots &0 \\
\dots &\dots&\dots&\dots&\dots&\dots&\dots&\dots \\
(-1)^{N}\lambda&0&\dots&0 &(-1)^{N-1}\frac{at^{0}_{1,0}t^{1}_{-1,0}}{t^{0}_{0,0}t^{1}_{0,0}}&(-1)^{N-2}\frac{at^{1}_{1,0}t^{2}_{-1,0}}{t^{1}_{0,0}t^{2}_{0,0}}&\dots &\frac{at^{N-1}_{1,0}t^{N}_{-1,0}}{t^{N-1}_{0,0}t^{N}_{0,0}}
\end{array} \right),
\end{equation*}
\begin{equation*} \label{DNg}
G=\left(\begin{array}{cccccccc}
1&0&\dots&\frac{b^{N+1}}{\lambda}\frac{t^{1}_{0,1}t^{N-1}_{0,0}}{t^{0}_{0,1}t^{N}_{0,0}} &\frac{ab}{\lambda}\frac{(t^{1}_{0,1})^2t^{1}_{-1,0}}{t^{0}_{0,1}t^{0}_{0,0}t^{1}_{0,0}}&\frac{ab^2}{\lambda}\frac{t^{1}_{0,1}t^{2}_{0,1}t^{2}_{-1,0}}{t^{0}_{0,1}t^{1}_{0,0}t^{2}_{0,0}}&\dots &\frac{ab^{N}}{\lambda}\frac{t^{1}_{0,1}t^{N}_{0,1}t^{N}_{-1,0}}{t^{0}_{0,1}t^{N-1}_{0,0}t^{N}_{0,0}} \\
\frac{bt^{2}_{0,1}t^{0}_{0,0}}{t^{1}_{0,0}t^{1}_{0,1}}&1&\dots&0&0&0&\dots &0 \\
\dots &\dots&\dots& \dots&\dots&\dots&\dots&\dots \\
0&\dots&\frac{bt^{N}_{0,1}t^{N-2}_{0,0}}{t^{N-1}_{0,0}t^{N-1}_{0,1}}&1&0&0&\dots &0 \\
0&\dots&0&b^N\frac{t^{0}_{0,0}t^{N-1}_{0,0}}{t^{1}_{-1,1}t^{N}_{0,0}}&\frac{at^{1}_{-1,0}t^{1}_{0,1}}{t^{1}_{0,0}t^{1}_{-1,1}}&ab\frac{t^{0}_{0,0}t^{2}_{-1,0}t^{2}_{0,1}}{t^{1}_{-1,1}t^{1}_{0,0}t^{2}_{0,0}}&\dots &ab^{N-1}\frac{t^{0}_{0,0}t^{N}_{-1,0}t^{N}_{0,1}}{t^{1}_{-1,1}t^{N-1}_{0,0}t^{N}_{0,0}} \\
0&\dots&0&b^{N-1}\frac{t^{1}_{0,0}t^{N-1}_{0,0}}{t^{2}_{-1,1}t^{N}_{0,0}}&0&\frac{at^{2}_{-1,0}t^{2}_{0,1}}{t^{2}_{0,0}t^{2}_{-1,1}}&\dots &ab^{N-2}\frac{t^{1}_{0,0}t^{N}_{-1,0}t^{N}_{0,1}}{t^{2}_{-1,1}t^{N-1}_{0,0}t^{N}_{0,0}} \\
\dots &\dots&\dots&\dots&\dots&\dots&\dots&\dots \\
0&\dots&0&b\frac{(t^{N-1}_{0,0})^2}{t^{N}_{-1,1}t^{N}_{0,0}}&0&0&\dots&a\frac{t^{N}_{-1,0}t^{N}_{0,1}}{t^{N}_{0,0}t^{N}_{-1,1}}
\end{array} \right).
\end{equation*}

In the particular case $N=2$ we have
\begin{eqnarray}\label{D23}
at^0_{0,0}t^0_{1,1}-t^0_{1,0}t^0_{0,1}=b(t^1_{0,1})^2,\nonumber\\
at^1_{0,0}t^1_{1,1}-t^1_{1,0}t^1_{0,1}=bt^0_{1,0}t^2_{0,1},\\
at^2_{0,0}t^2_{1,1}-t^2_{1,0}t^2_{0,1}=b(t^1_{1,0})^2.\nonumber
\end{eqnarray}
The Lax pair for (\ref{D23}) is of the form (\ref{LaxD2N1}) where $F$ and $G$ are $4\times 4$ matrices
\begin{eqnarray}\label{D23_F}
F=\left(
\begin{array}{cccc}
\frac{t^1_{1,0}t^0_{0,0}}{t^1_{0,0}t^0_{1,0}}&-1&0&0\\
-\lambda&\frac{t^{2}_{1,0}t^1_{0,0}}{t^{2}_{0,0}t^1_{1,0}}&
\frac{at^0_{1,0}t^1_{-1,0}}{t^0_{0,0}t^1_{0,0}}&-\frac{at^1_{1,0}t^2_{-1,0}}{t^1_{0,0}t^2_{0,0}}\\
-\lambda&0&\frac{at^0_{1,0}t^1_{-1,0}}{t^0_{0,0}t^1_{0,0}}&0\\
\lambda&0&-\frac{at^0_{1,0}t^1_{-1,0}}{t^0_{0,0}t^1_{0,0}}&\frac{at^1_{1,0}t^2_{-1,0}}{t^1_{0,0}t^2_{0,0}}
\end{array}\right),
\end{eqnarray}
\begin{eqnarray}\label{D23_G}
G=\left(
\begin{array}{cccc}
1&\frac{b^3}{\lambda}\frac{t^1_{0,1}t^1_{0,0}}{t^0_{0,1}t^2_{0,0}}&
\frac{ab}{\lambda}\frac{(t^1_{0,1})^2t^1_{-1,0}}{t^0_{0,0}t^0_{0,1}t^1_{0,0}}&
\frac{ab^2}{\lambda}\frac{t^1_{0,1}t^2_{-1,0}t^2_{0,1}}{t^0_{0,1}t^1_{0,0}t^2_{0,0}}\\
\frac{bt^0_{0,0}t^2_{0,1}}{t^1_{0,0}t^1_{0,1}}&1&0&0\\
0&\frac{b^2t^0_{0,0}t^1_{0,0}}{t^1_{-1,1}t^2_{0,0}}&\frac{at^1_{0,1}t^1_{-1,0}}{t^1_{0,0}t^1_{-1,1}}&
\frac{abt^0_{0,0}t^2_{-1,0}t^2_{0,1}}{t^1_{-1,1}t^1_{0,0}t^2_{0,0}}\\
0&\frac{b(t^1_{0,0})^2}{t^2_{-1,1}t^2_{0,0}}&0&
\frac{at^2_{-1,0}t^2_{0,1}}{t^2_{0,0}t^2_{-1,1}}
\end{array}\right).
\end{eqnarray}
Below in (\ref{LaxD23}) the system is written in a more familiar way. For the particular case $a=b=1$ the Lax pairs presented in this section have been found earlier in \cite{GHY}.

\section{Evaluation of the continuum limit}

Let us briefly discuss the continuum limit in the exponential type quad system (\ref{expon}) with arbitrary constant matrix $A$. To this end we assume that $a=1+o(\delta^2)$, $b=-\delta^2+o(\delta^2)$, when $\delta\rightarrow0$. We assume that smooth functions $v^j(x,y)$ exist such that $v^j(x,y)=u^j_{n,m}$, where $x=n\delta$, $y=m\delta$ and $1\leq j\leq N$. Then evidently we have
\begin{equation}\label{x}
u^j_{n+1,m}=v^j+\delta v_x^j+\frac{\delta^2}{2}v_{xx}^j+o(\delta^2),
\end{equation}
\begin{equation}\label{y}
u^j_{n,m+1}=v^j+\delta v_y^j+\frac{\delta^2}{2}v_{yy}^j+o(\delta^2),
\end{equation}
\begin{equation}\label{xy}
u^j_{n+1,m+1}=v^j+\delta(v^j_x+ v_y^j)+\frac{\delta^2}{2}(v_{xx}^j+2v_{xy}^j+ v_{yy}^j)+o(\delta^2).
\end{equation}
We substitute (\ref{x})-(\ref{xy}) into (\ref{expon}) and after some transformation we obtain a relation 
$v^{i}_{x,y}=\exp({\sum_{j=1}^N a_{ij}v^{j}})+o(1),\quad \mbox{for}\quad \delta\rightarrow0,\quad 1\leq i\leq N$
showing that quad system (\ref{expon}) goes in the continuum limit to an exponential type system in partial derivatives 
\begin{equation}\label{hypPDE} 
v^{i}_{x,y}=\exp({\sum_{j=1}^N a_{ij}v^{j}}),\quad 1\leq i\leq N.
\end{equation}
Due to the fact that the generalized Cartan matrix $A=\{a_{i,j}\}$ is degenerate, the system (\ref{hypPDE}) admits reducing of the order. For the reduced system the Lax representation is given in terms of the Cartan-Weyl basis in \cite{Leznov79, Drinfeld}.

Let us concentrate now on the continuum limit for the quad systems corresponding to $D^{(2)}_N$ at the level of the Lax pairs. Our formulas below differ from those used in \cite{Drinfeld}, since in \cite{Drinfeld} the coefficient matrix $\{a_{i,j}\}$ in the system (\ref{hypPDE}) denotes the Cartan transposed matrix. This leads to the fact that the system (\ref{hypPDE}), corresponding to the algebra $D^{(2)}_N$ in our work coincides with the system corresponding to the algebra $C^{(1)}_{N-1}$ in \cite{Drinfeld}. Up to this mismatch, the continuum limit completely coincides with the Drinfeld-Sokolov system.

Below we illustrate in detail the continuum limit for $D^{(2)}_3$, since in the general case $D^{(2)}_N$ it is evaluated in a similar way. Let us first change the variables $\Phi=\sigma\tilde\Phi$ in the system (\ref{D23_F}), (\ref{D23_G}). Here $\sigma=diag (1,\delta,\delta^3,\delta^2)$ is a diagonal matrix. We also replace $t^j_{n,m}=e^{-u^j_{n,m}}$. As a result we arrive at the system
\begin{equation}\label{tildaLax}
\tilde\Phi_{10}=\tilde F\tilde\Phi,\quad \tilde\Phi_{01}=\tilde G\tilde\Phi,
\end{equation}
where 
\begin{eqnarray*}\label{D23_tildeF}
\tilde{F}=\left(
\begin{array}{cccc}
e^{u^1-u^1_{1,0}+u^0_{1,0}-u^0}&-\delta&0&0\\
-\xi\delta^3&e^{u^2-u^2_{1,0}+u^1_{1,0}-u^1}&a\delta^2e^{u^0-u^0_{1,0}+u^1-u^1_{-1,0}}&-a\delta e^{u^1-u^1_{1,0}+u^2-u^2_{-1,0}}\\
-\xi\delta&0&ae^{u^0-u^0_{1,0}+u^1-u^1_{-1,0}}&0\\
\xi\delta^2&0&-a\delta e^{u^0-u^0_{1,0}+u^1-u^1_{-1,0}}&ae^{u^1-u^1_{1,0}+u^2-u^2_{-1,0}}
\end{array}\right),
\end{eqnarray*}
\begin{eqnarray*}\label{D23_tildeG}
\tilde{G}=\left(
\begin{array}{cccc}
1&-\frac{\delta^3}{\xi}e^{u^0_{0,1}-u^1_{0,1}-u^1+u^2}&-\frac{a\delta}{\xi}e^{u^0+u^0_{0,1}+u^1-2u^1_{0,1}-u^1_{-1,0}}&\frac{a\delta^2}{\xi}e^{u^0_{0,1}+u^1+u^2-u^1_{0,1}-u^2_{-1,0}-u^2_{0,1}}\\
-\delta e^{u^1-u^0+u^1_{0,1}-u^2_{0,1}}&1&0&0\\
0&\delta^2e^{u^1_{-1,1}-u^0-u^1+u^2}&ae^{u^1-u^1_{0,1}+u^1_{-1,1}-u^1_{-1,0}}&-a\delta e^{u^1_{-1,1}+u^1+u^2-u^0-u^2_{-1,0}-u^2_{0,1}}\\
0&-\delta e^{u^2_{-1,1}+u^2-2u^1}&0&ae^{u^2+u^2_{-1,1}-u^2_{-1,0}-u^2_{0,1}}
\end{array}\right).
\end{eqnarray*}

It is easily observed that due to the representations (\ref{x})-(\ref{xy}) potentials $\tilde F$ and $\tilde G$ admit asymptotic expansions of the form
\begin{equation}\label{deltagoestozero}
\tilde F=E+\delta\tilde F+o(\delta),\quad \tilde G=E+\delta\tilde G+o(\delta),\quad \mbox{for}\quad \delta\rightarrow0,
\end{equation}
where $E$ is the unity matrix.

We assume that eigenfunction $\tilde\Phi$ is represented as follows $\tilde\Phi_{n,m}=\Psi(x,y)$ where $x=n\delta$, $y=m\delta$ and $\Psi$ is a smooth function of the variables $x$ and $y$. Then we can write 
\begin{eqnarray}
&&\tilde\Phi_{n+1,m}=\Psi(x,y)+\delta\Psi(x,y)_x+o(\delta), \nonumber\\
&&\tilde\Phi_{n,m+1}=\Psi(x,y)+\delta\Psi(x,y)_y+o(\delta), \quad \delta\rightarrow0.\label{11}
\end{eqnarray}
Now evidently formulas (\ref{deltagoestozero}) and (\ref{11}) imply a system of the linear PDE
\begin{equation}\label{hatLax}
\Psi_{x}=R\Psi,\quad \Psi_{y}=S\Psi,
\end{equation}
where 
\begin{eqnarray*}
R=\left(
\begin{array}{cccc}
v^0_x-v^1_x&-1&0&0\\
0&v^1_x-v^2_x&0&-1\\
-\xi&0&v^1_x-v^0_x&0\\
0&0&-1&v^2_x-v^1_x
\end{array}\right),
\end{eqnarray*}
\begin{eqnarray*}
S=\left(
\begin{array}{cccc}
0&0&-\xi^{-1}e^{2v^0-2v^1}&0\\
-e^{2v^1-v^0-v^2}&0&0&0\\
0&0&0&-e^{2v^1-v^0-v^2}\\
0&-e^{-2v^1+2v^2}&0&0
\end{array}\right).
\end{eqnarray*}

The consistency condition of the system (\ref{hatLax}) leads to a system of the partial differential equations
\begin{eqnarray}\label{raznvxy}
\begin{array}{l}
v^0_{x,y}-v^1_{x,y}=e^{2v^0-2v^1}-e^{-v^0+2v^1-v^2},\\
v^1_{x,y}-v^2_{x,y}=e^{-v^0+2v^1-v^2}-e^{-2v^1+2v^2},
\end{array}
\end{eqnarray}
which doesn't coincide with the continuum limit of the quad system (\ref{D23}), having the form
\begin{eqnarray}\label{vxy}
\begin{array}{l}
v^0_{x,y}=e^{2v^0-2v^1},\\
v^1_{x,y}=e^{-v^0+2v^1-v^2},\\
v^2_{x,y}=e^{-2v^1+2v^2},
\end{array}
\end{eqnarray}
as might be expected by virtue of the formula (\ref{hypPDE}), but system (\ref{raznvxy}) can be rewritten as a reduction of (\ref{vxy}) obtained by introducing new variables $w^0=v^0-v^1$, $w^1=v^1-v^2$:
\begin{eqnarray}\label{wxy}
\begin{array}{l}
w^0_{x,y}=e^{2w^0}-e^{-w^0+w^1},\\
w^1_{x,y}=e^{-w^0+w^1}-e^{-2w^1}.
\end{array}
\end{eqnarray}
The latter belongs to the class of the generalized Toda lattices, studied in \cite{Drinfeld}. Under appropriate linear transformation the Lax pair (\ref{hatLax}) for the system (\ref{wxy}) is brought to the standard form \cite{Drinfeld, Leznov79}:
\begin{equation}\label{Laxeqw}
\phi_{x}=f\phi,\quad \phi_{y}=g\phi,
\end{equation}
where 
\begin{eqnarray*}
f=\left(
\begin{array}{cccc}
-w^0_x&0&0&-\zeta\\
-\zeta&-w^1_x&0&0\\
0&-\zeta&w^1_x&0\\
0&0&-\zeta&w^0_x
\end{array}\right),
\end{eqnarray*}
\begin{eqnarray*}
g=\left(
\begin{array}{cccc}
0&-\zeta^{-1}e^{w^1-w^0}&0&0\\
0&0&-\zeta^{-1}e^{-2w^1}&0\\
0&0&0&-\zeta^{-1}e^{w^1-w^0}\\
-\zeta^{-1}e^{2w^0}&0&0&0
\end{array}\right).
\end{eqnarray*}

\section{Formal asymptotics of the Lax operators eigenfunctions around singular values of $\lambda$ and local conservation laws of the quad systems}

The asymptotic behavior of the system of differential equations with respect to a parameter in the vicinity of the singular value of this parameter is an important characteristic of the system. A detailed presentation of the methods of studying these asymptotics can be found in Wasow's famous book \cite{Wasow}. In the theory of integrability, the mentioned asymptotics find applications in solving the scattering problem, in describing integrals of motion, and constructing symmetries of nonlinear equations (see \cite{Zakharov}). For the discrete equations with a parameter such a problem is rather difficult. Some particular cases are studied in \cite{H1985, HabYang, Mikhailov15}, which are not fit in the case of (\ref{LaxD2N1}). Hence we use here a suitable scheme suggested in \cite{HabKharxiv2019}. Below we briefly explain the algorithm.

Let us consider a system of the discrete linear equations
\begin{equation}  \label{lineardiscrete}
Y_{n+1}=f_nY_n, \, f_n=\sum^{\infty}_{j=-1}f_n^{(j)}\lambda^{-j},
\end{equation}
where $f_n^{(j)}\in{\bf C}^{k\times k}$ for $j\geq-1$ are matrix valued functions.
 In order to identify the matrix structure of the potential we divide the matrices into blocks as
\begin{equation} \label{A}
A=\left( \begin{array}{cc}
A_{11}&A_{12}\\
A_{21}&A_{22}
\end{array} \right),
\end{equation}
where the blocks $A_{11}$, $A_{22}$ are square matrices. Here we assume that in (\ref{lineardiscrete}) the coefficient $f_n^{(-1)}$ is of one of the forms
\begin{equation} \label{f_01}
f_n^{(-1)}=\left( \begin{array}{cc}
0&0\\
0&A_{22}
\end{array} \right),\quad \det A_{22}\neq 0,
\end{equation}
or 
\begin{equation} \label{f_02}
f_n^{(-1)}=\left( \begin{array}{cc}
A_{11}&0\\
0&0
\end{array} \right), \quad \det A_{11}\neq 0.
\end{equation}
Now our goal is to bring (\ref{lineardiscrete}) to a block-diagonal form 
\begin{equation}\label{diag}
\varphi_{n+1}=h_n\varphi_n
\end{equation}
where $h_n$ is a formal series
\begin{equation}\label{series1}
h_n=h_n^{(-1)}\lambda +  h_n^{(0)}+h_n^{(1)}\lambda^{-1}+h_n^{(2)}\lambda^{-2}+\cdots
\end{equation}
with the coefficients having the block structure
\begin{equation} \label{coeff-h}
h_n^{(j)}=\left( \begin{array}{cc}
(h_n^{(j)})_{11}&0\\
0&(h_n^{(j)})_{22}
\end{array} \right).
\end{equation}
To this end we use the linear transformation $Y_n=T_n\varphi_n$ assuming that $T_n$ is also a formal series
\begin{equation*}
T_n=E+T_n^{(1)}\lambda^{-1}+T_n^{(2)}\lambda^{-2}+\cdots, \label{series2}
\end{equation*}
where $E$ is the unity matrix and $T_n^{(j)}$ is a matrix with vanishing block-diagonal part:
\begin{equation*} \label{Tj1}
T_n^{(j)}=\left( \begin{array}{cc}
0&(T_n^{(j)})_{12}\\
(T_n^{(j)})_{21}&0
\end{array} \right).
\end{equation*}
After substitution of $Y_n=T_n\varphi_n$ into (\ref{lineardiscrete}) we get
\begin{equation}  \label{Th}
T_{n+1}h_n=\left(\sum^{\infty}_{j=-1}f_n^{(j)}\lambda^{-j}\right)T_n,
\end{equation}
where $h_n=\varphi_{n+1}\varphi^{-1}_n$.
Let us replace in (\ref{Th}) the factors by their formal expansions:
$$(E+T_{n+1}^{(1)}\lambda^{-1}+\cdots)(h_n^{(-1)}\lambda+  h_n^{(0)}+\cdots )=(f_n^{(-1)}\lambda+f_n^{(0)}+\cdots)(E+T_n^{(1)}\lambda^{-1}+\cdots) .$$
By comparing coefficients at the powers of $\lambda$ we derive a sequence of equations
\begin{eqnarray}\label{sequence1}
&&h_n^{(-1)}= f_n^{(-1)} \\
 \label{sequencek}
&&T_{n+1}^{(k)}h_n^{(-1)}+h_n^{(k-1)}-f_n^{(-1)} T_n^{(k)}=R^k_n,\quad k\geq1.
\end{eqnarray}
Here $R^k_n$ denotes terms that have already been found in the previous steps.

To find the unknown coefficients $T_n^{(j)}$, we must solve linear equations, that look like difference equations. However due to the special form of the coefficient $f_n^{(-1)}$ these equations are linear algebraic and therefore are solved without ``integration". In other words $T_n^{(j)}$ and $h_n^{(j)}$ are local functions of the potential since depend on a finite number of the shifts of the functions $f_n^{(-1)}$, $f_n^{(0)}$, $f_n^{(1)}$, etc. Indeed, equation (\ref{sequencek}) obviously implies 
\begin{equation*} \label{Tj2}
\left( \begin{array}{cc}
0&(T^{(k)}_{n+1})_{12}A_{22}\\
0&0
\end{array} \right)+\left( \begin{array}{cc}
p&0\\
0&q
\end{array} \right)-\left( \begin{array}{cc}
0&0\\
A_{22}(T^{(k)}_n)_{21}&0
\end{array} \right)=R^k_n
\end{equation*}
where $p=(h_n^{(k-1)})_{11}$, $q=(h_n^{(k-1)})_{22}$.
Evidently this equation is easily solved and the searched matrices $T_n^{(k)}$ and $h_n^{(k-1)}$ are uniquely found for any $k\geq 1$.

Suppose now that a system of equations of the form
\begin{equation}  \label{Y10Y01}
Y_{n+1,m}=(f_{n,m}^{(-1)}\lambda+f_{n,m}^{(0)}+\cdots)Y_{n,m}, \quad 
Y_{n,m+1}=G_{n,m}(\lambda)Y_{n,m},
\end{equation}
where $G_{n,m}(\lambda)$ is analytic at a vicinity of $\lambda=\infty$, is the Lax pair for the nonlinear quad system 
\begin{equation*}  \label{F}
F([u^{j}_{n,m}])=0,
\end{equation*}
i.e. $F$ depends on the variable $u^{j}_{n,m}$ and on its shifts with respect to the variables $j,n,m$.

Assume that function $f^{(-1)}_{n,m}$ has the structure (\ref{f_01}). Then due to the reasonings above there exists a linear transformation $Y_{n,m}\longmapsto \varphi_{n,m}=T_{n,m}^{-1}Y_{n,m}$ which reduces the first equation in (\ref{Y10Y01}) to a block-diagonal form (\ref{diag})-(\ref{coeff-h}). It can be checked that this transformation brings also the second equation of (\ref{f_01}) to an equation of the same block structure 
\begin{equation*}  \label{diagS}
\varphi_{n,m+1}=S_{n,m}\varphi_{n,m},
\end{equation*}
where $S_{n,m}$ is a formal power series 
\begin{equation*}\label{seriesS}
S_{n,m}=S_{n,m}^{(0)}+S_{n,m}^{(1)}\lambda^{-1}+S_{n,m}^{(2)}\lambda^{-2}+\cdots
\end{equation*}
and
\begin{equation*} \label{coeff-S}
S_{n,m}^{(j)}=\left( \begin{array}{cc}
(S_{n,m}^{(j)})_{1,1}&0\\
0&(S_{n,m}^{(j)})_{22}
\end{array} \right).
\end{equation*}
Since the compatibility property of linear systems is preserved under a change of the variables, we have the relation
\begin{equation*} \label{comp_hS}
S_{n+1,m}h_{n,m}=h_{n,m+1}S_{n,m}
\end{equation*}
which implies due to the block-diagonal structure that
\begin{equation} \label{cl_hS}
(D_n-1)\log \det (S)_{ii}=(D_m-1)\log \det (h)_{ii}, \, i=1,2.
\end{equation}
By evaluating and comparing the coefficients at the powers of $\lambda$ we derive the sequence of the local conservation laws.

The Lax pairs considered in the article have also the second singular point $\lambda=0$, so we briefly discuss the Lax pair represented as
\begin{equation}  \label{lambda0}
Y_{n+1,m}=F_{n,m}Y_{n,m}, \, 
Y_{n,m+1}=(g_{n,m}^{(-1)}\lambda^{-1}+g_{n,m}^{(0)}+g_{n,m}^{(1)}\lambda+\cdots)Y_{n,m},
\end{equation}
where $F_{n,m}=F_{n,m}(\lambda)$ is analytic at a vicinity of $\lambda=0$. We request that here the term $g_{n,m}^{(-1)}$  has the block structure (\ref{f_02}). In this case the block-diagonalization is performed in a way very similar to one recalled above for the point $\lambda=\infty$.

\subsection{System corresponding to affine Lie algebra $D^{(2)}_N$.}

Note that the procedure of formal diagonalization of the Lax pair provides an effective way to construct infinite series of the local conservation laws. Procedure of finding of the formal diagonalization of a linear system (\ref{lineardiscrete}) satisfying (\ref{f_01}) or (\ref{f_02}) is purely algorithmic. However, in order to apply the method to an arbitrary linear system we must transform it to an appropriate form and this step may lead to some difficulties. 

In the case related to $D^{(2)}_N$ we overcome these difficulties by applying the linear transformations 
\begin{equation*} \label{transPhiY}
\Phi=HY \quad (\mbox{or}\quad  \Phi=\bar HY)
\end{equation*}
converting the Lax equations (\ref{LaxD2N1}) to the suitable form (\ref{Y10Y01}) (or, respectively, (\ref{lambda0})), where the factor $H$ is lower (or, $\bar H$ is upper) block-triangular matrix  
\begin{equation} \label{H}
H=\left( \begin{array}{cc}
H_{11}&0\\
H_{21}&H_{22}
\end{array} \right),\quad \left(\mbox{or}\quad 
\bar H=\left( \begin{array}{cc}
\bar H_{11}&\bar H_{12}\\
0&\bar H_{22}
\end{array} \right)\right).
\end{equation}
Here we use the block representation (\ref{A}) where the blocks $A_{ij}$ are square matrices of the size $(N-1)\times(N-1)$. Note that (\ref{H}) $H$ and $\bar H$ have the same block structure. Moreover, the blocks $H_{11}$, $H_{22}$, $\bar H_{11}$ and $\bar H_{22}$ are diagonal matrices, some entries of which depend on the spectral parameter $\lambda$. Factors $H$ and $\bar H$ are effectively found (see, for example, (\ref{HD23}), (\ref{HmD23})). Below we illustrate all of the computations with the example.

{\bf Example 1.} Let us briefly discuss the quad system 
\begin{equation} \label{discD23}
\left\{ \begin{array}{c}
at^0_{0,0}t^0_{1,1}-t^0_{1,0}t^0_{0,1}=b(t^1_{0,1})^2,\\
at^1_{0,0}t^1_{1,1}-t^1_{1,0}t^1_{0,1}=bt^0_{1,0}t^2_{0,1},\\
at^2_{0,0}t^2_{1,1}-t^2_{1,0}t^2_{0,1}=b(t^1_{1,0})^2
\end{array} \right.
\end{equation}
corresponding to the algebra $D^{(2)}_3$. Its Lax pair reads as
\begin{equation}  \label{LaxD23}
\Phi_{1,0}=F\Phi, \quad  \Phi_{0,1}=G\Phi,
\end{equation}
where the potentials 
\begin{equation*} \label{fD23}
F=\left( \begin{array}{cccc}
\frac{t^0_{0,0}t^1_{1,0}}{t^0_{1,0}t^1_{0,0}}&-1&0&0\\
-\lambda&\frac{t^1_{0,0}t^2_{1,0}}{t^1_{1,0}t^2_{0,0}}&\frac{at^0_{1,0}}{t^0_{0,0}t^1_{0,0}}&-\frac{at^1_{1,0}}{t^1_{0,0}t^2_{0,0}}\\
-t^1_{0,0}\lambda&0&\frac{at^0_{1,0}}{t^0_{0,0}}&0\\
t^2_{0,0}\lambda&0&-\frac{at^0_{1,0}t^2_{0,0}}{t^0_{0,0}t^1_{0,0}}&\frac{at^1_{1,0}}{t^1_{0,0}}
\end{array} \right), \qquad 
G=\left( \begin{array}{cccc}
1&\frac{b^3t^1_{0,0}t^1_{0,1}}{\lambda t^0_{0,1}t^2_{0,0}}&\frac{ab(t^1_{0,1})^2}{\lambda t^0_{0,0}t^0_{0,1}t^1_{0,0}}&\frac{ab^2t^1_{0,1}t^2_{0,1}}{\lambda t^0_{0,1}t^1_{0,0}t^2_{0,0}}\\
\frac{bt^0_{0,0}t^2_{0,1}}{t^1_{0,0}t^1_{0,1}}&1&0&0\\
0&\frac{b^2t^0_{0,0}t^1_{0,0}}{t^2_{0,0}}&\frac{at^1_{0,1}}{t^1_{0,0}}&\frac{abt^0_{0,0}t^2_{0,1}}{t^1_{0,0}t^2_{0,0}}\\
0&\frac{b(t^1_{0,0})^2}{t^2_{0,0}}&0&\frac{at^2_{0,1}}{t^2_{0,0}}
\end{array} \right)
\end{equation*}
are not in a suitable form for the application of formal diagonalization. Therefore we have to do the transformation $\Phi=HY,$ with the block-triangular factor $H$:
\begin{equation} \label{HD23}
H=\left( \begin{array}{cccc}
1&0&0&0 \\ 
0&\xi&0&0 \\
\frac{(t^1_{-1,0})^2t^2_{0,0}}{t^1_{0,0}t^2_{-1,0}}+a\frac{t^1_{0,0}t^2_{-2,0}}{t^2_{-1,0}}&t^1_{-1,0}\xi&1&0\\
-\frac{t^1_{-1,0}t^2_{0,0}}{t^1_{0,0}}&-t^2_{-1,0}\xi&0&1
\end{array} \right), \quad \xi=\sqrt{\lambda}.
\end{equation}
The new variable $Y$ solves the system of equations
\begin{equation*}  \label{Y10Y01D23}
Y_{1,0}=fY, \quad 
Y_{0,1}=gY,
\end{equation*}
where the potential $g$ is analytic at $\xi=\infty$ and $f$ is given by 
\begin{equation*}
f=f^{(-1)}\xi+f^{(0)}+f^{(1)}\xi^{-1},
\end{equation*}
the factor $f^{(-1)}$ has the block-diagonal structure
\begin{equation*} \label{fm1D23}
f^{(-1)}=\left( \begin{array}{cccc}
0&-1&0&0\\ -1&0&0&0\\ 0&0&0&0\\ 0&0&0&0
\end{array} \right).
\end{equation*}
According to the reasoning above we can apply the formal diagonalization algorithm and find the series $T$, $h$ and $S$. Here we illustrate the first few coefficients 
\begin{eqnarray*}\label{hneqD23}
&&h=\left( \begin{array}{cccc} 0&-1&0&0 \\ -1&0&0&0 \\ 0&0&0&0 \\ 0&0&0&0 \end{array}\right)\xi+\left( \begin{array}{cccc} \frac{t^0_{0,0}t^1_{1,0}}{t^0_{1,0}t^1_{0,0}}&0&0&0 \\ 0&\frac{t^1_{0,0}t^2_{1,0}}{t^1_{1,0}t^2_{0,0}}+\frac{at^0_{1,0}t^1_{-1,0}}{t^0_{0,0}t^1_{0,0}}+\frac{at^1_{1,0}t^2_{-1,0}}{t^1_{0,0}t^2_{0,0}}&0&0 \\ 0&0&0&\frac{at^1_{1,0}}{t^2_{0,0}} \\ 0&0&0&0 \end{array}\right)+\nonumber\\
&&+\left( \begin{array}{cccc} 0&0&0&0 \\ \frac{at^0_{1,0}(t^1_{-1,0})^2t^2_{0,0}}{t^0_{0,0}(t^1_{0,0})^2t^2_{-1,0}}+\frac{a^2t^0_{1,0}t^2_{-2,0}}{t^0_{0,0}t^2_{-1,0}}+\frac{at^1_{-1,0}t^1_{1,0}}{(t^1_{0,0})^2}&0&0&0 \\ 0&0&0&0 \\ 0&0&0&0 \end{array}\right)\xi^{-1}+\nonumber\\
&&+\left( \begin{array}{cccc} 0&0&0&0 \\ 0&h^2_{22}&0&0 \\ 0&0&-\frac{at^2_{1,0}}{t^2_{0,0}}-\frac{a^2t^0_{1,0}t^1_{-1,0}t^1_{1,0}}{t^0_{0,0}(t^1_{0,0})^2}-\frac{a^2(t^1_{1,0})^2t^2_{-1,0}}{(t^1_{0,0})^2t^2_{0,0}}&\frac{at^0_{0,0}t^1_{1,0}t^2_{1,0}}{t^0_{1,0}(t^2_{0,0})^2}+\frac{a^2t^1_{-1,0}(t^1_{1,0})^2}{(t^1_{0,0})^2t^2_{0,0}}+\frac{a^2t^0_{0,0}(t^1_{1,0})^3t^2_{-1,0}}{t^0_{1,0}(t^1_{0,0}t^2_{0,0})^2} \\ 0&0& \frac{at^2_{1,0}}{t^1_{0,0}}&-\frac{at^0_{0,0}t^1_{1,0}t^2_{1,0}}{t^0_{1,0}t^1_{0,0}t^2_{0,0}} \end{array}\right)\xi^{-2}+\dots,
\end{eqnarray*}
where $h^2_{22}=\frac{at^0_{-1,0}t^1_{-1,0}t^0_{1,0}t^2_{0,0}}{(t^0_{0,0})^2t^1_{0,0}t^2_{-1,0}}+\frac{a^2t^0_{1,0}t^1_{-2,0}}{t^0_{0,0}t^1_{-1,0}}+\frac{a^2t^0_{-1,0}t^0_{1,0}t^1_{0,0}t^2_{-2,0}}{(t^0_{0,0})^2t^1_{-1,0}t^2_{-1,0}}+\frac{at^0_{-1,0}t^1_{1,0}}{t^0_{0,0}t^1_{0,0}}$,

\begin{eqnarray*}\label{TneqD23}
T=\left( \begin{array}{cccc} 1&0&0&0 \\ 0&1&0&0 \\ 0&0&1&0 \\ 0&0&0&1 \end{array}\right)+\left( \begin{array}{cccc} 0&0&0&0 \\ 0&0&0&0 \\ 0&\frac{t^0_{-1,0}t^1_{-1,0}t^2_{0,0}}{t^0_{0,0}t^2_{-1,0}}+\frac{at^1_{-2,0}t^1_{0,0}}{t^1_{-1,0}}+\frac{at^0_{-1,0}(t^1_{0,0})^2t^2_{-2,0}}{t^0_{0,0}t^1_{-1,0}t^2_{-1,0}}&0&0 \\ 0&-\frac{t^0_{-1,0}t^2_{0,0}}{t^0_{0,0}}&0&0 \end{array}\right)\xi^{-1}+\nonumber\\
+\left( \begin{array}{cccc} 0&0&\frac{at^0_{1,0}}{t^0_{0,0}t^1_{0,0}}&-\frac{at^1_{1,0}}{t^1_{0,0}t^2_{0,0}} \\ 0&0&0&0 \\ T^2_{31}&0&0&0 \\ -\frac{at^1_{-2,0}t^2_{0,0}}{t^1_{-1,0}}-\frac{t^0_{-1,0}t^1_{-1,0}(t^2_{0,0})^2}{t^0_{0,0}t^1_{0,0}t^2_{-1,0}}-\frac{at^0_{-1,0}t^2_{-2,0}t^1_{0,0}t^2_{0,0}}{t^0_{0,0}t^1_{-1,0}t^2_{-1,0}}&0&0&0 \end{array}\right)\xi^{-2}+\dots,
\end{eqnarray*}
where $T^2_{31}=\frac{at^0_{-2,0}t^1_{0,0}}{t^0_{-1,0}}+\frac{2at^1_{-2,0}t^2_{0,0}}{t^2_{-1,0}}+\frac{t^0_{-1,0}t^2_{0,0}((t^1_{-1,0})^2t^2_{0,0}+2a(t^1_{0,0})^2t^2_{-2,0})}{t^0_{0,0}t^1_{0,0}(t^2_{-1,0})^2}+\frac{a^2t^1_{0,0}(t^0_{0,0}t^1_{-2,0}t^2_{-1,0}+t^0_{-1,0}t^1_{0,0}t^2_{-2,0})^2}{t^0_{-1,0}t^0_{0,0}(t^1_{-1,0}t^2_{-1,0})^2}$,

\begin{eqnarray*}\label{SneqD23}
&&S=\left( \begin{array}{cccc} 1&0&0&0 \\ 0&1&0&0 \\ 0&0&\frac{at^1_{0,1}}{t^1_{0,0}}&\frac{abt^0_{0,0}t^2_{0,1}}{t^1_{0,0}t^2_{0,0}} \\ 0&0&0&\frac{at^2_{0,1}}{t^2_{0,0}} \end{array}\right)+
\left( \begin{array}{cccc} 0&\frac{abt^1_{-1,0}(t^1_{0,1})^2}{t^0_{0,0}t^1_{0,0}t^0_{0,1}}-\frac{ab^2t^2_{-1,0}t^1_{0,1}t^2_{0,1}}{t^0_{0,1}t^1_{0,0}t^2_{0,0}}+\frac{b^3t^1_{0,0}t^1_{0,1}}{t^0_{0,1}t^2_{0,0}}&0&0 \\ \frac{bt^0_{0,0}t^2_{0,1}}{t^1_{0,0}t^1_{0,1}}&0&0&0 \\ 0&0&0&0 \\ 0&0&0&0 \end{array}\right)\xi^{-1}\nonumber\\
&&+\left( \begin{array}{cccc} \frac{abt^1_{0,1}}{t^0_{0,1}}\left(\frac{t^1_{0,1}(t^1_{-1,0})^2t^2_{0,0}}{t^0_{0,0}(t^1_{0,0})^2t^2_{-1,0}}+\frac{at^1_{0,1}t^2_{-2,0}}{t^0_{0,0}t^2_{-1,0}}-\frac{bt^1_{-1,0}t^2_{0,1}}{(t^1_{0,0})^2}\right)&0&0&0 \\ 0&0&0&0 \\ 0&0&S^2_{33}&S^2_{34} \\ 0&0&S^2_{43}&S^2_{44} \end{array}\right)\xi^{-2}+\dots,
\end{eqnarray*}
where 
$S^2_{33}=\frac{abt^1_{0,1}}{t^1_{0,0}t^0_{0,1}}\left(\frac{b^3t^0_{0,0}t^2_{0,1}}{t^2_{0,0}}-\frac{(at^1_{0,1})^2t^2_{-2,0}}{t^0_{0,0}t^2_{-1,0}}-\frac{a(bt^0_{0,0}t^2_{-1,0}t^2_{0,1}-t^1_{-1,0}t^1_{0,1}t^2_{0,0})^2}{(t^1_{0,0})^2t^0_{0,0}t^2_{-1,0}t^2_{0,0}}\right)$, \\
$S^2_{34}=\frac{ab^2t^2_{0,1}}{t^1_{0,0}t^0_{0,1}}\left(\frac{b^3t^2_{0,1}(t^0_{0,0})^2}{(t^2_{0,0})^2}+\frac{2abt^2_{0,1}t^1_{0,1}t^1_{-1,0}t^0_{0,0}}{(t^1_{0,0})^2t^2_{0,0}}-\frac{a(t^1_{0,1}t^1_{-1,0})^2}{(t^1_{0,0})^2t^2_{-1,0}}-\frac{a^2(t^1_{0,1})^2t^2_{-2,0}}{t^2_{-1,0}t^2_{0,0}}-\frac{ab^2t^2_{-1,0}(t^0_{0,0}t^2_{0,1})^2}{(t^1_{0,0}t^2_{0,0})^2}\right)$, \\
$S^2_{43}=\frac{abt^1_{0,1}t^2_{0,1}}{t^0_{0,1}}\left(\frac{at^1_{0,1}t^1_{-1,0}}{(t^1_{0,0})^2t^0_{0,0}}+\frac{b^2}{t^2_{0,0}}-\frac{abt^2_{0,1}t^2_{-1,0}}{(t^1_{0,0})^2t^2_{0,0}}\right)$,\\
$S^2_{44}=\frac{ab^2(t^2_{0,1})^2}{t^0_{0,1}t^2_{0,0}}\left(\frac{at^1_{0,1}t^1_{-1,0}}{(t^1_{0,0})^2}+\frac{b^2t^0_{0,0}}{t^2_{0,0}}-\frac{abt^2_{0,1}t^0_{0,0}t^2_{-1,0}}{t^2_{0,0}(t^1_{0,0})^2}\right).$

By virtue of the formulas (\ref{cl_hS}) we derive local conservation laws:
\begin{enumerate}
	\item[1.] $\left(D_n-1\right)\left(\frac{b^4t^0_{0,0}t^2_{0,1}}{t^0_{0,1}t^2_{0,0}}-\frac{a^2b(t^1_{0,1})^2t^2_{-2,0}}{t^0_{0,0}t^0_{0,1}t^2_{-1,0}}-\frac{ab(bt^0_{0,0}t^2_{0,1}t^2_{-1,0}-t^1_{-1,0}t^1_{0,1}t^2_{0,0})^2}{t^0_{0,0}t^0_{0,1}(t^1_{0,0})^2t^2_{-1,0}t^2_{0,0}}\right)=$
	\item[] $\left(D_m-1\right)\left(\frac{t^0_{0,0}t^2_{1,0}}{t^0_{1,0}t^2_{0,0}}+\frac{a^2t^0_{1,0}t^2_{-2,0}}{t^0_{0,0}t^2_{-1,0}}+\frac{a(t^0_{0,0}t^1_{1,0}t^2_{-1,0}+t^0_{1,0}t^1_{-1,0}t^2_{0,0})^2}{t^0_{0,0}t^0_{1,0}(t^1_{0,0})^2t^2_{-1,0}t^2_{0,0}}\right)$,
	\item[2.] $\left(D_n-1\right)\left(\frac{b^2}{2(t^0_{0,1})^2}\left(\frac{a^2(t^1_{0,1})^2t^2_{-2,0}}{t^0_{0,0}t^2_{-1,0}}-\frac{b^3t^0_{0,0}t^2_{0,1}}{t^2_{0,0}}+\frac{a(bt^0_{0,0}t^2_{-1,0}t^2_{0,1}-t^1_{-1,0}t^1_{0,1}t^2_{0,0})^2}{t^0_{0,0}(t^1_{0,0})^2t^2_{-1,0}t^2_{0,0}}\right)^2-\frac{a^2bt^0_{-2,0}(t^1_{0,1})^2}{t^0_{0,0}t^0_{0,1}t^0_{-1,0}}\right.$
	\item[] $\left.-\frac{ab}{t^0_{-1,0}t^0_{0,1}}\left(\frac{at^1_{0,1}t^1_{-2,0}}{t^1_{-1,0}}+\frac{t^0_{-1,0}t^1_{-1,0}t^1_{0,1}t^2_{0,0}}{t^0_{0,0}t^1_{0,0}t^2_{-1,0}}+\frac{at^0_{-1,0}t^1_{0,0}t^1_{0,1}t^2_{-2,0}}{t^0_{0,0}t^1_{-1,0}t^2_{-1,0}}-\frac{bt^0_{-1,0}t^2_{0,1}}{t^1_{0,0}}\right)^2\right)=$
	\item[] $\left(D_m-1\right)\left(\frac{a^2t^0_{1,0}t^0_{-2,0}}{t^0_{0,0}t^0_{-1,0}}+\frac{1}{2}\left(\frac{t^0_{0,0}t^2_{1,0}}{t^0_{1,0}t^2_{0,0}}+\frac{a^2t^0_{1,0}t^2_{-2,0}}{t^0_{0,0}t^2_{-1,0}}+\frac{a(t^0_{0,0}t^1_{1,0}t^2_{-1,0}+t^0_{1,0}t^1_{-1,0}t^2_{0,0})^2}{t^0_{0,0}t^0_{1,0}(t^1_{0,0})^2t^2_{-1,0}t^2_{0,0}}\right)^2\right.$
	\item[] $\left.\frac{a}{t^0_{-1,0}t^0_{1,0}}\left(\frac{t^0_{1,0}t^0_{-1,0}t^1_{-1,0}t^2_{0,0}}{t^0_{0,0}t^1_{0,0}t^2_{-1,0}}+\frac{at^0_{1,0}t^1_{-2,0}}{t^1_{-1,0}}+\frac{at^0_{-1,0}t^0_{1,0}t^1_{0,0}t^2_{-2,0}}{t^0t^1_{-1,0}t^2_{-1,0}}+\frac{t^1_{1,0}t^0_{-1,0}}{t^1_{0,0}}\right)^2\right)$,
	\item[3.] $\left(D_n-1\right)\left(-\frac{abt^0_{2,0}}{t^0_{0,0}t^0_{1,0}t^0_{0,1}}\left(\frac{at^1_{-1,0}t^1_{0,1}t^2_{0,0}-abt^0_{0,0}t^2_{-1,0}t^2_{0,1}+b^2t^0_{0,0}(t^1_{0,0})^2}{t^1_{0,0}t^2_{0,0}}\right)^2\right.$
	\item[] $\left.	-\frac{b}{t^0_{1,0}t^0_{0,1}}\left(\frac{a^2t^0_{1,0}t^1_{0,1}t^2_{-2,0}}{t^0_{0,0}t^2_{-1,0}}+\frac{at^0_{1,0}t^1_{0,1}(t^1_{-1,0})^2t^2_{0,0}}{t^0_{0,0}(t^1_{0,0})^2t^2_{-1,0}}+\frac{at^1_{1,0}t^1_{0,1}t^1_{-1,0}}{(t^1_{0,0})^2}-\frac{abt^0_{1,0}t^1_{-1,0}t^2_{0,1}}{(t^1_{0,0})^2}-\frac{abt^0_{0,0}t^1_{1,0}t^2_{-1,0}t^2_{0,1}}{(t^1_{0,0})^2t^2_{0,0}}+\frac{b^2t^0_{0,0}t^1_{1,0}}{t^2_{0,0}}\right)^2\right.$
	\item[] $\left.-\frac{b^2}{2(t^0_{0,1})^2}\left(\frac{b^3t^0_{0,0}t^2_{0,1}}{t^2_{0,0}}-\frac{a^2(t^1_{0,1})^2t^2_{-2,0}}{t^0t^2_{-1,0}}-\frac{a(bt^0_{0,0}t^2_{0,1}t^2_{-1,0}-t^1_{0,1}t^1_{-1,0}t^2_{0,0})^2}{t^0_{0,0}(t^1_{0,0})^2t^2_{-1,0}t^2_{0,0}}\right)^2\right)=$
	\item[] $\left(D_m-1\right)\left(\frac{a^2t^0_{0,0}t^0_{3,0}}{t^0_{1,0}t^0_{2,0}}+\frac{at^0_{0,0}(t^1_{2,0})^2}{(t^1_{1,0})^2t^0_{2,0}}+\frac{at^0_{2,0}}{t^0_{0,0}}\left(\frac{t^0_{0,0}(t^1_{0,0})^2t^2_{1,0}+at^0_{1,0}t^1_{-1,0}t^1_{1,0}t^2_{0,0}+at^0_{0,0}(t^1_{1,0})^2t^2_{-1,0}}{t^0_{1,0}t^1_{0,0}t^1_{1,0}t^2_{0,0}}\right)^2\right.$
	\item[] $\left.+\frac{2at^1_{2,0}\left(t^0_{0,0}(t^1_{0,0})^2t^2_{1,0}+at^0_{1,0}t^1_{-1,0}t^1_{1,0}t^2_{0,0}+at^0_{0,0}(t^1_{1,0})^2t^2_{-1,0}\right)}{t^0_{1,0}t^1_{0,0}(t^1_{1,0})^2t^2_{0,0}}+\frac{1}{2}\left(\frac{t^0_{0,0}t^2_{1,0}}{t^0_{1,0}t^2_{0,0}}+\frac{a^2t^0_{1,0}t^2_{-2,0}}{t^0_{0,0}t^2_{-1,0}}+\frac{a(t^0_{0,0}t^1_{1,0}t^2_{-1,0}+t^0_{1,0}t^1_{-1,0}t^2_{0,0})^2}{t^0_{0,0}t^0_{1,0}(t^1_{0,0})^2t^2_{-1,0}t^2_{0,0}}\right)^2\right)$.
\end{enumerate}

In a similar way we investigate the system around the point $\lambda=0$. To this end we first change the variables, $\bar{\Phi}=\bar{H}Y$ where
\begin{equation} \label{HmD23}
\bar{H}=\left( \begin{array}{cccc}
\xi^{-1}&0&0&0\\ 0&1&-\frac{a}{b^2}\frac{t^1_{0,1}t^2_{0,0}}{t^0_{0,0}(t^1_{0,0})^2}&-\frac{a}{b}\frac{t^2_{0,1}}{(t^1_{0,0})^2}\\ 0&0&1&0\\ 0&0&0&1
\end{array} \right), 
\end{equation}
that reduces the system to the suitable form
\begin{equation*}  \label{Y10Y01mD23}
Y_{1,0}=\bar{f}(\xi)Y, \quad 
Y_{0,1}=\left(\bar{g}^{(-1)}\xi^{-1}+\bar{g}^{(0)}\right)Y,
\end{equation*}
where $\bar{f}(\xi)$ is analytic at the vicinity of $\xi=0$ and the matrix $\bar{g}^{(-1)}$ has the appropriate block-diagonal structure:
\begin{equation*} \label{gm1mD23}
\bar{g}^{(-1)}=\left( \begin{array}{cccc}
0&\frac{b^3t^1_{0,0}t^1_{0,1}}{t^0_{0,1}t^2_{0,0}}&0&0\\ 
\frac{bt^0_{0,0}t^2_{0,1}}{t^1_{0,1}t^1_{0,0}}&0&0&0\\ 
0&0&0&0\\ 0&0&0&0
\end{array} \right).
\end{equation*}
Therefore one can perform the diagonalization procedure and find the local conservation laws:
\begin{enumerate}
	\item[1.] $\left(D_m-1\right)\left(-b\frac{(t^1_{0,1})^2t^2_{0,0}+a(t^1_{0,0})^2t^2_{0,2}}{t^0_{0,0}t^0_{1,0}t^2_{0,1}}\right)=$
	\item[] $\left(D_n-1\right)\left(\frac{a^2t^0_{0,0}t^2_{0,3}}{t^0_{0,1}t^2_{0,2}}+\frac{t^0_{0,1}t^2_{0,0}}{t^0_{0,0}t^2_{0,1}}+\frac{2at^1_{0,0}t^1_{0,2}}{(t^1_{0,1})^2}+\frac{at^0_{0,0}(t^1_{0,2})^2t^2_{0,1}}{t^0_{0,1}(t^1_{0,1})^2t^2_{0,2}}+\frac{at^0_{0,1}(t^1_{0,0})^2t^2_{0,2}}{t^0_{0,0}(t^1_{0,1})^2t^2_{0,1}}\right)$,
	\item[2.] $\left(D_m-1\right)\left(\frac{b^2t^1_{0,1}t^2_{0,0}}{t^0_{0,0}t^1_{1,0}}-\frac{abt^2_{0,0}t^0_{0,-1}t^0_{1,0}t^1_{0,1}}{t^0_{0,0}t^1_{1,0}(t^1_{0,0})^2}-\frac{abt^0_{1,0}t^1_{0,-1}t^2_{0,1}}{(t^1_{0,0})^2t^1_{1,0}}\right)=$
	\item[] $\left(D_n-1\right)\left(\frac{a^2t^0_{0,-1}t^2_{0,2}}{t^0_{0,0}t^2_{0,1}}+\frac{at^1_{0,0}t^1_{0,2}}{(t^1_{0,1})^2}+\frac{at^0_{0,-1}(t^1_{0,1})^2t^2_{0,0}}{t^0_{0,0}t^2_{0,1}(t^1_{0,0})^2}+\frac{t^0_{0,1}t^2_{0,0}}{t^0_{0,0}t^2_{0,1}}+\frac{a(t^1_{0,0})^2t^0_{0,1}t^2_{0,2}}{(t^1_{0,1})^2t^0_{0,0}t^2_{0,1}}+\frac{at^1_{0,1}t^1_{0,-1}}{(t^1_{0,0})^2}\right)$,
	\item[3.] $\left(D_m-1\right)\left(-\frac{abt^0_{0,2}(t^1_{0,0})^2}{t^0_{0,0}t^0_{0,1}t^0_{1,0}}-\frac{b^2}{2}\left(\frac{t^2_{0,0}(t^1_{0,1})^2+a(t^1_{0,0})^2t^2_{0,2}}{t^0_{0,0}t^0_{1,0}t^2_{0,1}}\right)^2-\frac{b}{t^0_{1,0}t^0_{0,1}}\left(\frac{t^0_{0,1}(t^1_{0,1})^2t^2_{0,0}+at^0_{0,0}t^1_{0,0}t^1_{0,2}t^2_{0,1}+at^0_{0,1}(t^1_{0,0})^2t^2_{0,2}}{t^0_{0,0}t^1_{0,1}t^2_{0,1}}\right)^2\right)=$
	\item[] $\left(D_n-1\right)\left(\frac{a^2t^0_{0,0}t^0_{0,3}}{t^0_{0,1}t^0_{0,2}}+\frac{a^3t^0_{0,0}(t^1_{0,3})^2}{t^0_{0,2}(t^1_{0,2})^2}+\frac{at^0_{0,2}}{t^0_{0,0}}\left(\frac{at^0_{0,0}(t^1_{0,1})^2t^2_{0,3}+t^1_{0,0}t^1_{0,2}t^0_{0,1}t^2_{0,2}+t^0_{0,0}(t^1_{0,2})^2t^2_{0,1}}{t^0_{0,1}t^1_{0,1}t^1_{0,2}t^2_{0,2}}\right)^2\right.$
	\item[] $\left.+\frac{2a^2t^1_{0,3}(at^0(t^1_{0,1})^2t^2_{0,3}+t^0_{0,1}t^1_{0,0}t^1_{0,2}t^2_{0,2}+t^0_{0,0}(t^1_{0,2})^2t^2_{0,1})}{t^0_{0,1}t^1_{0,1}(t^1_{0,2})^2t^2_{0,2}}+\frac{1}{2}\left(\frac{a^2t^0_{0,0}t^2_{0,3}}{t^0_{0,1}t^2_{0,2}}+\frac{t^0_{0,1}t^2_{0,0}}{t^0_{0,0}t^2_{0,1}}+\frac{a(t^0_{0,0}t^1_{0,2}t^2_{0,1}+t^0_{0,1}t^1_{0,0}t^2_{0,2})^2}{t^0_{0,0}t^0_{0,1}(t^1_{0,1})^2t^2_{0,1}t^2_{0,2}}\right)^2\right)$.
\end{enumerate}

\section{Higher symmetries}
Quad systems (\ref{D2-N}) and (\ref{A1-N}) possess higher symmetries. However, presented in the variables $t^j_{nm}$ the symmetries have non-localities. They become local in the potential variables introduced as follows $r^j=\frac{t^j_{1,0}}{t^j_{0,0}}$.

Let us concentrate on the system (\ref{discD23}). By setting  $u=\frac{t^1_{1,0}}{t^1_{0,0}}$, $v=\frac{t^2_{1,0}}{t^2_{0,0}}$ and $w=\frac{t^3_{1,0}}{t^3_{0,0}}$ we convert it to the form
\begin{equation} \label{x2}
\left. \begin{array}{l}
au_{1,1}=u_{1,0}+v_{0,1}^2\left(\frac{a}{u}-\frac{1}{u_{0,1}}\right),\\
av_{1,1}=v_{1,0}+u_{1,0}w_{0,1}\left(\frac{a}{v}-\frac{1}{v_{0,1}}\right),\\
aw_{1,1}=w_{1,0}+v_{1,0}^2\left(\frac{a}{w}-\frac{1}{w_{0,1}}\right).\\
\end{array} \right.
\end{equation}

It is easily checked that a system of the linear equations
\begin{equation}  \label{x3}
\phi_{1,0}=f\phi, \quad  \phi_{0,1}=g\phi,
\end{equation}
with
\begin{equation*} \label{ffD23}
f=\left( \begin{array}{cccc}
\frac{v}{u}&-1&0&0\\
-\lambda&\frac{w}{v}&au&-av\\
-\frac{\lambda}{v}&0&\frac{au}{v}&0\\
\frac{\lambda}{w}&0&-\frac{au}{w}&\frac{av}{w}
\end{array} \right), 
\end{equation*}
\begin{equation*} \label{ggD23} 
g=\left( \begin{array}{cccc}
1&\frac{(au_{0,1}-u)(av_{0,1}-v)(aw_{0,1}-w)}{\lambda uv^2}&\frac{a(au_{0,1}-u)}{\lambda}&\frac{a(au_{0,1}-u)(av_{0,1}-v)}{\lambda u}\\
\frac{av_{0,1}-v}{u}&1&0&0\\
0&\frac{(av_{0,1}-v)(aw_{0,1}-w)}{uv^2}&a&\frac{a(av_{0,1}-v)}{u}\\
0&\frac{aw_{0,1}-w}{v^2}&0&a
\end{array} \right)
\end{equation*}
provides the Lax pair for quad system (\ref{x2}).

Apparently quad system  (\ref{x2}) possesses a hierarchy of higher symmetries. Here we represent the simplest of them
\begin{equation} \label{symteqD23}
\left. \begin{array}{l}
u_{t}=w+\frac{2auv}{v_{-1,0}}+\frac{av^2}{w_{-1,0}}+  \frac{au^2w_{-1,0}}{v_{-1,0}^2}+\frac{a^2u^2}{w_{-2,0}},\\
v_t=av_{1,0}+\frac{au_{1,0}w}{v}+ \frac{vw}{u}   +\frac{av^2}{v_{-1,0}}  +\frac{a^2u_{1,0}v}{w_{-1,0}} +\frac{av^3}{uw_{-1,0}} ,\\
w_{t}=a^2u_{2,0}+\frac{w^2}{u}+\frac{2av_{1,0}w}{v}+\frac{au_{1,0}w^2}{v^2}+\frac{av^2_{1,0}}{u_{1,0}}.
\end{array} \right.
\end{equation}
Besides, it obviously has classical symmetries
$u_t=u,\quad v_t=v,\quad w_t=w$
and
$u_t=(-1)^nu,\quad v_t=0,\quad w_t=(-1)^nw.$

Symmetry (\ref{symteqD23}) admits the Lax pair $\varphi_{1,0}=f\varphi$, $\varphi_{t}=A\varphi$, where $f$ is given in (\ref{x3}) and $A$ is as follows
\begin{equation*} \label{AD23} 
A=\left( \begin{array}{cccc}
\frac{a^2u}{w_{-2,0}}+\frac{av}{v_{-1,0}}+\frac{auw_{-1,0}}{(v_{-1,0})^2}&\frac{au}{v_{-1,0}}+\frac{av}{w_{-1,0}}&-au&av\\
0&\frac{av}{v_{-1,0}}+\frac{av^2}{uw_{-1,0}}+\frac{a^2u_{1,0}}{w_{-1,0}}&-av&a^2u_{1,0}+\frac{av^2}{u}\\
\left(\frac{w_{-1,0}}{(v_{-1,0})^2}+\frac{a}{w_{-2,0}}\right)\lambda&\frac{1}{v_{-1,0}}\lambda&-\frac{av}{v_{-1,0}}-\frac{a^2u}{w_{-2,0}}-\frac{auw_{-1,0}}{(v_{-1,0})^2}-\lambda&0\\
-\frac{1}{v_{-1,0}}\lambda&-\frac{1}{w_{-1,0}}\lambda&\frac{au}{v_{-1,0}}+\frac{av}{w_{-1,0}}&-\lambda-\frac{av}{v_{-1,0}}-\frac{av^2}{uw_{-1,0}}-\frac{a^2u_{1,0}}{w_{-1,0}}
\end{array} \right).
\end{equation*}

\section{Conclusions}

In the article the problem of integrable discretization of the generalized two-dimensional Toda lattices is discussed. This kind of the lattices have appeared in $18$-th century in the frame of the Laplace cascade integration method of hyperbolic type linear PDE. The lattices have applications in the field theory, geometry, integrability theory etc. (see \cite{Darboux, Toda, Mikhailov, Leznov, ShabatYamilov, Olshanetsky, LeznovSmirnovShabat, Kac}).

Nowadays various classes of the discrete versions of the Toda lattices are known \cite{Doliwa, Willox, Xenitidis, Fu, Nijhoff, Kuniba}. They are intensively studied due to the applications in the discrete field theories \cite{Kuniba}.

In the present article we studied discrete systems on the quadrilateral graph corresponding to the series of the affine Lie algebras $D^{(2)}_N$ of the form (\ref{expon}),
which is a generalization of that suggested earlier in \cite{GHY}. We discussed in detail the algorithm of constructing the Lax pairs for these systems and showed also that the Lax pairs allow to find infinite series of the local conservation laws. In the continuum limit the systems convert to the systems of partial differential equations  studied in \cite{Drinfeld}.



\begin{thebibliography}{0,0}

\bibitem{AdlerStartsev}
V.E. Adler and S.Ya. Startsev, On discrete analogues of the Liouville equation, \textit{Theor. Math. Phys.}, \textbf{121} (1999) 1484--1495.

\bibitem{Caudrelier2018}  
J. Avan, V. Caudrelier and N. Cramp$\acute{e}$, From Hamiltonian to zero curvature formulation for classical integrable boundary conditions \textit{J. Phys. A: Math. Theor.} \textbf{51} (2018) 30LT01.

%
\bibitem{Biondini2009}  
G. Biondini and G. Hwang, Solitons, boundary value problems and a nonlinear method of images, \textit{J. Phys. A: Math. Theor.} \textbf{42} (2009) 205207.

%
\bibitem{Caudrelier2014}  
V. Caudrelier and Q.C. Zhang, Yang--Baxter and reflection maps from vector solitons with a boundary, \textit{Nonlinearity} \textbf{27} (2014) 1081.

\bibitem{Darboux} 
G. Darboux, Lecons sur la th\'eorie g\'en\'erale des surfaces et les applications geometriques du calcul infinitesimal, \textit{Paris:
Gauthier-Villars} \textbf{1--4} (1896) 513~p., 579~p., 512~p., 547~p.

%
\bibitem{Doliwa} 
A. Doliwa, Non-commutative lattice-modified Gel'fand-Dikii systems, \textit{J. Phys. A: Math. Theor.} \textbf{46} (2013) 205202.

%
\bibitem{Drinfeld} 
V.G. Drinfeld and V.V.  Sokolov, Lie algebras and equation of KdV type, \textit{J. Sov. Math.} \textbf{30} (1985) 1975--2036.

\bibitem{Xenitidis} 
A.P. Fordy and P. Xenitidis, Zn graded discrete Lax pairs and integrable difference equations, \textit{J. Phys. A: Math. Theor.} \textbf{50} (2017)
165205.

\bibitem{Kac} 
I.B. Frenkel and V.G. Kac, Basic Representations of Affine Lie Algebras and Dual Resonance Models, \textit{Inv. math.} \textbf{62} (1980) 23--66.

%
\bibitem{Fu} 
W. Fu, Direct linearisation of the discrete-time two-dimensional Toda lattices, \textit{J. Phys. A: Math. Theor.} \textbf{51} (2018) 334001.

%
\bibitem{GHY} 
R.N.  Garifullin, I.T. Habibullin and M.V. Yangubaeva, Affine and finite Lie algebras and integrable Toda field equations on discrete space-time, \textit{ SIGMA} \textbf{8} (2012) 33 pp. 

%
\bibitem{H1985}  
I.T. Habibullin, The discrete Zakharov-Shabat system and integrable equations, (Russian) Translated in J. Soviet Math. \textbf{40} (1988) 108-115. Differential geometry, Lie groups and mechanics. VII. Zap. Nauchn. Sem. Leningrad. Otdel. Mat. Inst. Steklov. (LOMI) \textbf{146} (1985), 137--146, 203, 207.

%
\bibitem{Habibullin1990} 
I.T. Habibullin, Backlund transformation and integrable boundary-initial value problems, \textit{Nonlinear world} \textbf{1} (1990) 130--138.

%
\bibitem{Habibullin2005} 
I.T. Habibullin, Truncations of Toda chains and the reduction problem, \textit{Theor. and Math. Phys.} \textbf{143} 2005 515--528.  

%
\bibitem{HabYang} 
I.T. Habibullin and M.V. Yangubaeva, Formal diagonalization of a discrete Lax operator and conservation laws and symmetries of dynamical systems,  \textit{Theor. Math. Phys.} \textbf{177} (2013) 1655--1679.

%
\bibitem{HabKharxiv2019} 
I.T. Habibullin and A.R. Khakimova, Discrete exponential type systems on a quad graph, corresponding to the affine Lie algebras $A^{(1)}_{N-1}$, 	arXiv:1901.03486 [nlin.SI] (2019).

%
\bibitem{Hirota} 
R. Hirota, Discrete analogue of a generalized Toda equation, \textit{J. Phys. Soc. Jpn.} \textbf{50} (1981) 3785--3791.

%
\bibitem{Krichever}  
I. Krichever, O. Lipan, P. Wiegmann and A. Zabrodin, Quantum integrable systems and elliptic solutions of classical discrete nonlinear equations, \textit{Commun. Math. Phys.} \textbf{188} (1997) 267--304. 

%
\bibitem{Kuniba} 
A. Kuniba, T. Nakanishi and J. Suzuki, T-systems and Y-systems in integrable systems, \textit{J. Phys. A: Math. Theor.} \textbf{44} (2011) 103001 146 pp.

%
\bibitem{Leznov79}
A. N. Leznov and M. V. Saveliev, Representation of zero curvature for the system of nonlinear partial differential equations Xa,zz = exp (KX)u and its integrability, \textit{Lett. Math. Phys.} \textbf{3} (1979) 489--494.

%
\bibitem{Leznov} 
A.N. Leznov, On the complete integrability of a nonlinear system of partial differential equations in two-dimensional space, \textit{Theor. Math. Phys.} \textbf{42} (1980) 225--229.

%
\bibitem{LeznovSmirnovShabat} 
A.N. Leznov, V.G. Smirnov and A.B. Shabat, The group of internal symmetries and the conditions of integrability of two-dimensional dynamical systems, \textit{Theor. Math. Phys.} \textbf{51} (1982) 322--330.

%
\bibitem{Mikhailov} 
A.V. Mikhailov, Integrability of a two-dimensional generalization of the Toda chain, \textit{JETP Lett.} \textbf{30} (1979) 414--418.

%
\bibitem{Olshanetsky} 
A.V. Mikhailov, M.A. Olshanetsky and A.M. Perelomov, Two-dimensional generalized Toda lattice, \textit{Commun. Math. Phys.} \textbf{79} (1981) 473--488.

%
\bibitem{Mikhailov15} 
A.V. Mikhailov, Formal diagonalisation of Lax-Darboux schemes, \textit{Model. Anal. Inform. Sist.} \textbf{22} (2015) 795--817.

%
\bibitem{Miwa} 
T. Miwa, On Hirota's difference equations, \textit{Proc. Japan Acad.} \textbf{58A} (1982) 9--12.

%
\bibitem{Nijhoff} 
F.W. Nijhoff, V.G. Papageorgiou, H.W. Capel and G.R.W. Quispel, The lattice Gel'fand-Dikii hierarchy, \textit{Inverse Problems} \textbf{8}  (1992) 597421.

%
\bibitem{Novikov}
S. Novikov and I. Dynnikov, Discrete spectral symmetries of low-dimensional differential operators and difference operators on regular lattices and two-dimensional manifolds, \textit{Russian Mathematical Surveys} \textbf{52} (1997) 175--234.

%
\bibitem{ShabatYamilov} 
A.B. Shabat and R.I. Yamilov, Exponential systems of type I and Cartan matrices, \textit{Preprint, OFM BFAN SSSR, Ufa} (1981).

%
\bibitem{Sklyanin} 
E.K. Sklyanin, Boundary conditions for integrable equations, \textit{Funct. Anal. Appl.} \textbf{21} (1987) 164–-166.

%
\bibitem{Smirnov} 
S.V. Smirnov, \textit{Theor. Math. Phys.} \textbf{182} (2015) 189--210.

%
\bibitem{Toda} 
M. Toda, Vibration of a Chain with Nonlinear Interaction, \textit{J. Phys. Soc. Japan} \textbf{22} (1967) 431--436.

%
\bibitem{Wasow} 
W. Wasow, Asymptotic Expansions for Ordinary Differential Equations, \textit{Dover Books on Advanced Mathematics (Dover: Dover Pubns)}  (1987) 374 pp.

%
\bibitem{Willox} 
R. Willox and M. Hattori, Discretisations of Constrained KP Hierarchies, \textit{J. Math. Sci. Univ. Tokyo} \textbf{22} (2015) 613--661.

%
\bibitem{Wilson} 
G. Wilson, The modified Lax and two-dimensional Toda lattice equations associated with simple Lie algebras, \textit{Ergod. Theory Dyn. Syst.} \textbf{1} (1981) 361--380.

%
\bibitem{ZabrodinJETP} 
A. Zabrodin, Hidden quantum R-matrix in discrete time classical Heisenberg magnet, \textit{Preprint ITEP-TH-54/97, Inst. Theor. Exp. Phys., Moscow, solv-int/9710015, JETP Letters} \textbf{66} (1997) 653--659.

%
\bibitem{ZabrodinTMF} 
A.V. Zabrodin, Hirota's difference equations, \textit{Theor. Math. Phys.} \textbf{113} (1997) 1347--1392. 

%
\bibitem{Zakharov} 
V.E. Zakharov, S.V. Manakov, S.P. Novikov and L.P. Pitaevskii, Theory of Solitons: Method of the Inverse Problem [in Russian] (Moscow: Nauka 1980).


\end{thebibliography}
\end{document}